\documentclass{article}
\usepackage{amsfonts}
\usepackage{amsmath}
\usepackage{amssymb}
\usepackage{amsthm}

\newtheorem{theorem}{Theorem}
\newtheorem{definition}{Definition}
\newtheorem{proposition}{Proposition}

\newcommand{\bi}{\begin{itemize}}
\newcommand{\ei}{\end{itemize}}
\newcommand{\be}{\begin{equation}}
\newcommand{\ee}{\end{equation}}
\newcommand{\ben}{\begin{enumerate}[(i)]}
\newcommand{\een}{\end{enumerate}}

\newcommand{\LL}{\ensuremath{\mathcal{L}}}

\newcommand{\KK}{\ensuremath{\mathcal{K}}}

\newcommand{\NN}{\mathbb{N}}

\newcommand{\bh}{\backslash}

\newcommand{\parties}[1]{\ensuremath{\mathcal{P}(#1)}}

\newcommand{\belun}{\mathcal{B}_1}
\newcommand{\pbelun}{\mathcal{B}_1^p}
\newcommand{\beldeux}{\mathcal{B}_2}
\newcommand{\pbeldeux}{\mathcal{B}_2^p}

\DeclareMathOperator{\init}{init}
\DeclareMathOperator{\update}{update}

\DeclareMathOperator{\act}{act}

\newcommand{\distrib}[1]{\mathcal{D}(#1)}

\newcommand{\prob}[3]{\mathbb{P}^{#1}_{#2}\left({#3}\right)}

\newcommand{\states}{K}

\newcommand{\actionsun}{I}
\newcommand{\actionsdeux}{J}
\newcommand{\signauxun}{C}
\newcommand{\signauxdeux}{D}
\newcommand{\trans}{p}

\newcommand{\targets}{T}

\newcommand{\ini}{\delta}
\newcommand{\tp}[2]{p(#1\mid #2)}

\title{Determinacy and Decidability of Reachability\\ Games with Partial Observation on Both Sides}
\author{Nathalie Bertrand, Blaise Genest, Hugo Gimbert}

\begin{document}

\maketitle
\begin{abstract}
We consider two-players stochastic reachability games
with partial observation on both sides and finitely many states, signals and actions.
We prove that in such games, either player $1$ has a strategy for winning with probability $1$,
or player $2$ has such a strategy, or both players have strategies that guarantee
winning with non-zero probability (positively winning strategies).
We give a fix-point algorithm for deciding which of the three cases holds,
which can be decided in doubly-exponential time.
\end{abstract}

\section*{Introduction}
We prove two determinacy and decidability results about two-players stochastic reachability games
with partial observation on both sides and finitely many states, signals and actions.
Player $1$ wants the play to reach the set of target states,
while player $2$ wants to keep away the play from target states.
Players take their decisions based upon \emph{signals} that they receive all along
the play, but they cannot observe the actual state of the game, nor the actions played by their opponent,
nor the signals received by their opponent. Each player only observes the signals he receives
and the actions he plays.
Players have common knowledge of the initial state of the game.

Our determinacy result is of a special kind, it concerns two notions of solutions for stochastic games.
The first one is the well known notion of \emph{almost-surely} winning strategy, which guarantees winning with probability $1$
against any strategy of the opponent.
The second one is the notion of \emph{positively winning} strategy:
a strategy is positively winning if it guarantees a non-zero winning probability against any strategy of the opponent.
This notion is less known, to our knowledge it appeared recently in~\cite{theseflorian}.
The notion of positively winning strategy is different from the notion of positive value,
because the non-zero winning probability can be made arbitrarily small
by the opponent, hence existence of a positively winning strategy does not give any clue
for deciding whether the value is zero or not.
Existence of a positively winning strategy guarantees that the opponent does not have an almost-surely winning strategy,
however there is no straightforward reason that one of these cases should always holds. Actually,
if we consider more complex classes of games than reachability games,
there are various examples where neither player $1$ has a positively winning strategy
nor player $2$ has an almost-surely
winning strategy.

Our first result (Theorem~\ref{theo:twoplayers1}) states that, in reachability games with partial observation on both sides,
either player $1$ has a positively winning strategy or player $2$ has an almost-surely winning strategy.
Moreover which case holds is decidable in \emph{exponential} time.
Notice that an almost-surely winning strategy for player $2$ in a reachability game
is \emph{surely winning} as well.

Our second result (Theorem~\ref{theo:twoplayers2}) states that
either player $1$ has an almost-surely winning strategy
or player $2$ has a positively winning strategy,
and this is decidable in \emph{doubly-exponential} time.

Both these results strengthen and generalize in several ways results given in~\cite{chdr07}.
Actually, in this paper is addressed only the particular case where
player $2$ has perfect information
and target states are observable by player $1$.
Moreover in~\cite{chdr07} no determinacy result is established,
the paper "only" describes an algorithm for deciding whether player $1$ has an almost-sure winning strategy.

\section{Reachability games with partial observation on both sides}

We consider zero-sum stochastic games with partial observation on both sides,
where the goal of Player $1$ is to reach a certain set of target states.
Players only partially observe the state of the game,
via signals.
Signals and state transitions are governed by probability transitions:
when the state is $k$ and two actions $i$ and $j$ are chosen,
player $1$ and $2$ receive respectively signals $c$ and $d$ and the new state is $l$
with probability $\tp{c,d,l}{k,i,j}$.\\

\subsection{Notations}
We use the following standard notations~\cite{renault}.\\
The game is played in steps. At each step the game is in some state $k\in\states$.
The goal of player $1$ is to reach target states $\targets\subseteq\states$.
Before the game starts,
the initial state is chosen according to the initial distribution $\ini \in\distrib{\states}$,
which is common knowledge of both players.
Players $1$ and $2$ choose actions $i\in\actionsun$ and $j\in\actionsdeux$,
then player $1$ receives a signal $c\in\signauxun$,
player $2$ receives a signal $d\in\signauxdeux$,
and the game moves to a new state $l$.
This happens with probability $\tp{c,d,l}{k,i,j}$
given by fixed transition probabilities $\trans : K\times I \times J \to \distrib{C\times D \times K}$,
known by both players.
We denote $\tp{l}{k,i,j}=\sum_{c,d}\tp{c,d,l}{k,i,j}$.
Players observe and remember their own actions and the signals they receive,
it is convenient to suppose that in the signal they receive is encoded the action they just played,
formally their exists $\act : \signauxun \cup \signauxdeux \to \actionsun\cup\actionsdeux$ such that
$\tp{c,d,k'}{k,i,j}>0 \iff (i=\act(c) \text{ and } j=\act(d))$.
We denote
$\tp{c,d,l}{k}=\tp{c,d,l}{k,\act(i),\act(j)}$.
This way, plays can be described by sequences of states and signals for both players,
without mentioning which actions were played.
A sequence $p=(k_0,c_1,d_1,\ldots,c_{n},d_{n},k_{n})\in(\states\signauxun\signauxdeux)^*\states$ is a finite play
if for every $0\leq m< n$, $\tp{c_{m+1},d_{m+1},k_{m+1}}{k_m,\act(c_{m+1}),\act(d_{m+1})}>0$.
An infinite play is a sequence $p\in(\states\signauxun\signauxdeux)^\omega$ whose prefixes are finite plays.

A strategy of player $1$ is a mapping $\sigma : \distrib{\states}\times\signauxun^*\to \distrib{\actionsun}$
and a strategy of player $2$ is $\tau : \distrib{\states}\times\signauxdeux^*\to \distrib{\actionsdeux}$.

In the usual way, an initial distribution $\ini$ and two strategies $\sigma$ and $\tau$
define a probability measure $\prob{\sigma,\tau}{\ini}{\cdot}$ on the set of infinite plays,
equipped with the $\sigma$-algebra generated by cylinders.

We use random variables $K_n,I_n,J_n,C_n,D_n$ for designing respectively
the $n$-th state, action of player $1$, action of player $2$, signal of player $1$,
signal of player $2$.
The probability to reach a target state someday is:
\[
\gamma_1(\ini,\sigma,\tau) = \prob{\sigma,\tau}{\ini}{\exists m \in \NN, K_m \in\targets}\enspace,
\]
and the probability to never reach the target is $\gamma_2(\ini,\sigma,\tau)=1-\gamma_2(\ini,\sigma,\tau)$.
Player $1$ seeks maximizing $\gamma_1$ while player $2$ seeks maximizing $\gamma_2$.


\subsection{Winning almost-surely or positively}


\begin{definition}[Almost-surely and positively winning]
A distribution $\delta$
is \emph{almost-surely winning} for player $1$ if there exists a strategy $\sigma$
such that
\begin{equation}\label{eq:as}
\forall \tau,
\gamma_1(\ini,\sigma,\tau)=1
\enspace.
\end{equation}
A distribution $\delta$
is \emph{positively winning} for player $1$ if there exists a strategy $\sigma$
such that
\begin{equation}\label{eq:ps}
\forall \tau,
\gamma_1(\ini,\sigma,\tau)>0
\enspace.
\end{equation}
If the uniform distribution on a set of states $L\subseteq \states$
is almost-surely or positively winning then $L$
itself is said to be almost-surely or positively winning.
If there exists $\sigma$ such that~\eqref{eq:as}
holds for every almost-surely winning
distribution then $\sigma$ is said to be almost-surely winning
.

Positively winning strategies for player $1$ and almost-sure winning and positively winning strategies for player $2$ are defined
similarly.
\end{definition}

\section{Winning almost-surely and positively with finite memory}

Of special algorithmic interest are strategies with finite memory.
\begin{definition}[Strategies with finite memory]
A strategy $\sigma$ with finite memory is described by
a finite set $M$ called the memory,
a strategic function $\sigma_M:M\to\distrib{\actionsun}$,
an update function $\update_M : M \times \signauxun \to M$,
an initialization function $\init_M : \parties{\states} \to M$.

For playing with $\sigma$, player $1$ proceeds as follows.
Let $\delta$ be the initial distribution with support $L$,
then initially player $1$ puts the memory in state $\init_M(L)$.
When the memory is in state $m$,
player $1$ chooses his action according to the distribution
$\sigma_M(m)$.
When player $1$ receives a signal $c$ and its memory state
is $m$, he changes the memory state to $\update_M(m,c)$. 
\end{definition}




A crucial tool for establishing decidability and determinacy result is the class of finite memory strategy
whose finite memory if based on the notions of beliefs or pessimistic beliefs.

\subsection{Beliefs and pessimistic beliefs}

The belief of a player at some moment of the play is the set of states
he thinks the game could possibly be,
according to the signals he received up to now.
The pessimistic belief is similar,
except the player assumes that no final state has been reached yet.
One of the motivations for
introducing beliefs and pessimistic beliefs is Proposition~\ref{prop:belief}.

Beliefs of player $1$ are defined by mean of the operator $\belun$ that associates with $L\subseteq \states$
and $c\in\signauxun$,
\begin{equation}
\label{eq:defbelief}
\belun(L,c) = \{
k\in \states \mid \exists l\in L, \exists d\in \signauxdeux, \tp{k,c,d}{l}>0
\}\enspace.
\end{equation}
We defined inductively the belief after signals $c_1,\ldots,c_n$ by
$\belun(L,c_1,\ldots,c_n,c) = \belun(\belun(L,c_1,\ldots,c_n),c)$.

Pessimistic beliefs of player $1$ are defined by
\[
\pbelun(L,c) =\belun(L\bh \targets,c)\enspace.
\]

Beliefs $\beldeux$ and pessimistic beliefs $\pbeldeux$ for player $2$ are defined similarly.
We will use the following properties of beliefs and pessimistic beliefs.

\begin{proposition}\label{prop:belief}
Let $\sigma,\tau$ be strategies for player $1$ and $2$
and $\delta$ an initial distribution with support $L$.
Then for every $n\in\NN$,
\begin{align*}
&\prob{\sigma,\tau}{\delta}{K_{n+1}\in\belun(L,C_1,\ldots,C_n)}=1\enspace,\\
&\prob{\sigma,\tau}{\delta}{K_{n+1}\in\beldeux(L,D_1,\ldots,D_n)}=1\enspace,\\
&\prob{\sigma,\tau}{\delta}{K_{n+1}\in\pbelun(L,C_1,\ldots,C_n)\text{ or }K_m\in\targets \text{ for some }1\leq m\leq n}=1\enspace,\\
&\prob{\sigma,\tau}{\delta}{K_{n+1}\in\pbeldeux(L,D_1,\ldots,D_n)\text{ or }K_m\in\targets \text{ for some }1\leq m\leq n}=1\enspace.
\end{align*}

Suppose $\tau$ and $\delta$ almost-surely winning for player $2$, then for every $n\in\NN$,
\[
\prob{\sigma,\tau}{\delta}{\beldeux(L,D_1,\ldots,D_n)\text{ is a.s.w. for player }2}=1\enspace.
\]

Suppose $\sigma$ and $\delta$ almost surely winning for player $1$, then for every $n\in\NN$,
\[
\prob{\sigma,\tau}{\delta}{\pbelun(L,C_1,\ldots,C_n)\text{ is a.s.w. for player $1$ or } \exists 1\leq m\leq n, K_m\in\targets}=1\enspace.
\]
\end{proposition}

\begin{proof}
Almost straightforward from the definitions.
\end{proof}

\subsection{Belief and pessimistic belief strategies}

A strategy $\sigma$ is said to be a \emph{belief strategy} for player $1$ if it has finite memory $M = \parties{\states}$ and
\begin{enumerate}
\item the initial state of the memory is the support of the initial distribution,
\item the update function is $(L,c)\to \belun(L,c)$,
\item the strategic function $\parties{\states}\to\distrib{\actionsun}$ associates with each
memory state $L\subseteq \states$ the uniform distribution on a non-empty
set of actions $I_L\subseteq \actionsun$.
\end{enumerate}
The definition of a pessimistic belief strategy for player $1$ is the same, except the update function is $\pbelun$.

\section{Determinacy and decidability results}

In this section, we establish our main result,
a determinacy result of a new kind.
Usual determinacy results in game theory concern the existence of a value.
Here the determinacy refers to positive and almost-sure winning:

\begin{theorem}[Determinacy]\label{theo:determinacy}
Every initial distribution is either almost-surely winning for player $1$,
surely winning for player $2$ or positively winning for both players. 
\end{theorem}

Theorem~\ref{theo:determinacy} is a corollary of Theorems~\ref{theo:twoplayers1} and~\ref{theo:twoplayers2},
in which details are given about the complexity of deciding whether an initial distribution
is positively winning for player $1$ and whether it is positively winning for player $1$.

Deciding whether a distribution is positively winning for player $1$ is quite easy,
because player $1$ has a very simple strategy for winning positively: playing randomly any action.

\begin{theorem}[Deciding positive winning for player $1$]\label{theo:twoplayers1}
Every initial distribution is either positively winning for player $1$ or surely winning for player $2$.

The strategy for player $1$ which plays randomly any action is positively winning.
Player $2$ has a belief strategy which is surely winning.

The partition of supports between those positively winning for player $1$
and those surely winning for player $2$ is computable in time exponential in $|\states|$,
together with an almost-surely winning belief strategy for player $2$.
\end{theorem}

\begin{proof}[Proof of Theorem~\ref{theo:twoplayers1}]
Let $\LL_\infty\subseteq \parties{\states\bh\targets}$
be the greatest fix-point of the monotonic operator
$\Phi:\parties{\parties{\states\bh\targets}}\to \parties{\parties{\states\bh\targets}}$ defined by:
\[
\Phi(\LL)= \{L\in \LL \mid \exists j\in \actionsdeux, \forall d\in\signauxdeux, \text{ if }j=\act(d)\text{ then }\beldeux(L,d)\in \LL\},
\]
and let $\sigma_R$ be the strategy for player $1$ that plays randomly any action.
To establish Theorem~\ref{theo:twoplayers1} we are going to prove that:
\begin{enumerate}
\item[(A)] every support in $\LL_\infty$ is surely winning for player $2$, and
\item[(B)] $\sigma_R$ is positively winning from any support $L\subseteq\states$ which is not in $\LL_\infty$.
\end{enumerate}

We start with proving (A).
For winning surely from any support $L\in\LL_\infty$, player $2$ uses the following belief strategy: if the current belief of player $2$ is $L\in\LL_\infty$ then player $2$
chooses an action $j_L$ such that whatever signal $d$ player $2$ receives (with $\act(d)=j_L$),
his next belief $\beldeux(L,d)$ will be in $\LL_\infty$ as well.
By definition of $\Phi$ there always exists such an action $j$,
and this defines a belief-strategy $\sigma:L\to j_L$ for player $2$.
When playing with this strategy, beliefs of player $2$ never intersect $\targets$
hence according to Proposition~\ref{prop:belief},
against any strategy $\sigma$ of player $1$,
the play stays almost-surely in $\states\bh\targets$,
hence it stays surely in $\states\bh\targets$.

Conversely, we prove (B).
We fix the strategy for player $1$
which consists in playing randomly
any action with equal probability,
and
the game is a one-player game where only player $2$ has choices to make:
it is enough to prove (B)
in the special case where the set of actions of player $1$ is a singleton $\actionsun=\{i\}$.
Let $\LL_0=\parties{\states\bh\targets}\supseteq \LL_1=\Phi(\LL_0)\supseteq \LL_2=\Phi(\LL_1)\ldots$ and $\LL_\infty$ be the limit of this sequence, the greatest fixpoint of $\Phi$.
We prove that for any support $L\in\parties{\states}$, if $L\not\in\LL_\infty$ then:
\be\label{eq:postoprove}
\text{$L$ is positively winning for player $1$}\enspace.
\ee
If $L\cap\targets \not=\emptyset$,~\eqref{eq:postoprove} is obvious.
For delaing with the case where $L\in\parties{\states\bh\targets}$, 
we define for every $n\in\NN$, $\KK_n = \parties{\states\bh\targets} \bh \LL_n$,
and we prove by induction on $n\in\NN$
that for every $L\in\KK_n$,
then for every initial distribution $\delta_L$ with support $L$,
for every strategy $\tau$,
\be\label{eq:topo}
\prob{\tau}{\delta_L}{\exists m\in\NN, K_m\in\targets, 2\leq m\leq n+1 }>0 \enspace.
\ee
For $n=0$,~\eqref{eq:topo} is obvious because $\KK_0=\emptyset$.
Suppose that for some $n\in\NN$, \eqref{eq:topo} holds for every $L\in\KK_n$,
and let $L\in\KK_{n+1}$.
If $L\in\KK_{n}$ then by inductive hypothesis,~\eqref{eq:topo} holds.
Otherwise by definition of $\KK_{n+1}$, $L\in\LL_{n}\bh\Phi(\LL_n)$ hence by definition of $\Phi$,
whatever action $j$ is played by player $2$ at the first round,
there exists a signal $d_j$ such that $\act(d_j)=j$ and $\beldeux(L,d_j)\not \in \LL_n$.
Let $\tau$ be a strategy for player $2$ and $j$ an action such that $\tau(\delta_L)(j)>0$.
If $\beldeux(L,d_j)\cap\targets \not= \emptyset$ then according to Proposition~\ref{prop:belief},
$\prob{\tau}{\delta_L}{K_2\in\targets}>0$.
Otherwise $\beldeux(L,d_j)\in\parties{\states\bh\targets}\bh\LL_n=\KK_n$ hence according to the inductive hypothesis
$\prob{\tau[d_j]}{\beldeux(L,d_j)}{\exists m\in\NN, 2\leq m\leq n+1, K_m\in\targets}>0$.
Since player $1$ has only one action, by definition of beliefs,
for every state $l\in\beldeux(L_d,j)$, $\prob{\tau}{\delta_L}{K_2 =l}>0$.
Together with the previous equation, we obtain\\ $\prob{\tau}{\delta_L}{\exists m\in\NN, 3\leq m\leq n+2, K_m\in\targets}>0$.
This achieves the inductive step.

The computation of the partition of supports between those positively winning for player $1$,
and those surely winning for player $2$ and a surely winning strategy for player
$2$ amounts to the computation of the largest fixpoint of $\Phi$.
since $\Phi$ is monotonic, and each application of the operator
can be computed in exponential time,
the overall computation can be achieved in exponential time and space.
\end{proof}

Deciding whether an initial distribution is positively winning for player $1$ is easy because
player $1$ has a very simple strategy for that: playing randomly.
Player $2$ does not have such a simple strategy for winning positively:
he has to make hypotheses about the beliefs of player $1$,
as is shown in the example depicted by fig.~\ref{fig:blaise}.

\begin{figure}

\caption{\label{fig:blaise} A game where player $2$ needs a lot of memory.}
\end{figure}

\begin{theorem}[Deciding positive winning for player $2$]\label{theo:twoplayers2}
Every initial distribution is either almost-surely winning for player $1$ or positively winning for player $2$.

Player $1$ has an almost-surely winning strategy which is pessimistic belief.
Player $2$ has a finite memory strategy such that each memory
state is a pair of a state and a pessimistic belief of player $1$.

The partition of supports between those almost-surely winning for player $1$
and those positively winning for player $2$ is computable in time doubly-exponential in $|\states|$,
together with the winning strategies for both players.
\end{theorem}

The proof of Theorem~\ref{theo:twoplayers2} is based on the following intuition.
The easiest way of winning for player $2$ is to reach with positive probability a state
from where he wins surely. Hence player $1$ will try to prevent the play from reaching such
surely winning states, in other words player $1$ should prevent his pessimistic belief to contain such surely winning states.
However, doing so, player $1$ may prevent the play to reach a target state:
it may hold that player $2$ has a strategy for winning positively under the hypothesis that pessimistic beliefs of player $1$
never contains surely winning states. This adds new beliefs of player $1$ to the collection
of pessimistic beliefs that player $1$ should avoid. And so on...

For formalizing these intuitions, we make use of \emph{$\LL$-games}.
\begin{definition}[$\LL$-games]
Let $\LL\subseteq\states$ be a collection of supports.
The $\LL$-game associated with $\LL$ is the game with same actions, transitions and signals than the original
partial observation game, only the winning condition changes:
player $1$ loses if either the play never reaches a target state or
if at some moment the pessimistic belief of player $1$ is in $\LL$
and the play has never visited a target state previously.
Formally given an initial distribution $\delta$ with support $L$
and two strategies $\sigma$ and $\tau$ the winning probability of player $1$ is:
\[
\prob{\sigma,\tau}{\delta}{\exists n\geq 1, K_n\in\targets \text{ and }\forall m<n, \belun(L,C_1,\ldots,C_m)\not\in\LL}\enspace.
\]
\end{definition}

Actually $\LL$-games are special cases of reachability games,
as shown in the next proposition and its proof.
\begin{proposition}\label{prop:LLgames}
In a $\LL$-game, every support is either positively winning for player $2$
or almost-surely winning for player $1$.
This partition can be computed in time doubly-exponential in $|\states|$.
Player $2$ has a positively winning strategy whose states are pairs
of states and pessimistic beliefs of player $1$.
Player $1$ has an almost-surely winning pessimistic-belief strategy.
\end{proposition}

\begin{proof}
We define a reachability game $G_\LL$ associated with $\LL$ in the following way.
Make the synchronized product of the original game with pessimistic beliefs of player $1$:
each state is a pair $(k,L)$ with $k\in\states$ and $L\subseteq\states\bh\targets$.
Transitions are inherited from the original game
except that every state whose second component is in $\LL$ is absorbing.
The set of target states is the set of pairs whose first component is in $\targets$.
According to Theorem~\ref{theo:twoplayers1} in the reachability game $G_\LL$
every state is either positively winning
for player $2$ or almost-surely winning for player $1$.
Moreover according to Theorem~\ref{theo:twoplayers1},
player $2$ has a positively winning belief strategy $\tau_\LL$ in $G_\LL$ from which it is easy 
to construct a positively winning strategy in the $\LL$-game,
with finite memory, whose memory states are sets of states of $G_\LL$. 
Also according to Theorem~\ref{theo:twoplayers1},
player $1$ has an almost-surely pessimistic belief strategy $\sigma_\LL$ in $G_\LL$.
Notice that pessimistic beliefs of player $1$ in $G_\LL$ cannot take all the possible values
in $\parties{(\states\bh\targets)\times\parties{\states\bh\targets}}$ because intuitively
player $1$ has perfect knowledge about his own pessimistic beliefs and formally
such a pessimistic belief is always of the type $\cup_{l\in L} \{(l,L)\}$ for some $L\subseteq\states\bh\targets$.
As a consequence, it is easy to extract from $\sigma_\LL$ a pessimistic belief almost-surely winning strategy in
the $\LL$-game.
\end{proof}

The heart of the proof of Theorem~\ref{theo:twoplayers2} is based on the two next propositions.
\begin{proposition}\label{prop:upward}
Let $\LL$ be an upward-closed collection of supports.
Suppose that every support in $\LL$ is positively winning for player $2$
in the original game.

Then any support positively winning for player $2$ in the $\LL$-game is positively
winning in the original game as well.

If, apart from supports in $\LL$, there are no supports positively winning for player $2$ in the $\LL$-game,
then every support $L\not\in\LL$ is almost-surely winning for player $1$ in the original game.
\end{proposition}

\begin{proof}
Let $\LL_p$ be the set of supports positively winning in the $\LL$-game,
that are not in $\LL$.

We start with the case where $\LL_p$ is not empty.
Let $\tau$ be a strategy for player $2$ positively winning in the original game.
Let $\tau'$ be a strategy for player $2$ positively winning in the $\LL$-game.
Let $\tau''$ be the following strategy for player $2$.
Player $2$ starts playing totally randomly any action with equal probability.
At each step of the play, player $2$ throws a dice with three sides
to decide whether he should:
\begin{itemize}
\item keep playing randomly,
\item
pick randomly a support $L\in\LL$,
forget the past observations and 
switch definitively to strategy $\tau$
with initial support $L$,
\item
pick randomly a support $L\in\LL_p$,
forget the past observations and 
switch definitively to strategy $\tau'$
with initial support $L$.
\end{itemize}
Let us prove that $\tau''$ is positively winning in the original game,
i.e. for every strategy $\sigma$ and initial distribution $\delta$ with support $L\in\LL_p$,
\be\label{eq:taupp}
\prob{\sigma,\tau''}{\delta}{\exists n\geq 1, K_n\in\targets}<1\enspace.
\ee
By definition of $\tau''$,
there is non-zero probability that the play is consistent with $\tau'$ i.e.
\be\label{eq:taup}
\prob{\sigma,\tau''}{\delta}{\forall n\geq 1, J_n = \tau'(L,D_1,\ldots,D_{n-1})}>0\enspace.
\ee
Since $\tau'$ is positively winning in the $\LL$-game,
\be\label{eq:tauppp}
\prob{\sigma,\tau'}{\delta}{\exists n\geq 1, K_n\in\targets \text{ and }\forall m<n, \belun(L,C_1,\ldots,C_m)\not\in\LL}<1\enspace.
\ee
If $\prob{\sigma,\tau'}{\delta}{\exists n\geq 1, K_n\in\targets}<1$ then together with~\eqref{eq:taup} this gives~\eqref{eq:taupp}.\\
If $\prob{\sigma,\tau'}{\delta}{\exists n\geq 1, K_n\in\targets}=1$ then according to~\eqref{eq:tauppp}, there exists $N\geq 1$
and a pessimistic belief $B\in\LL$ such that:
\[
\prob{\sigma,\tau'}{\delta}{\belun(L,C_1,\ldots,C_N)=B\text{ and } \forall 1\leq m\leq N, K_m\not\in\targets}>0\enspace.
\]
Since every sequence of actions is played with positive probability by $\tau''$, then:
\be\label{eq:BBB}
\forall l\in B, \prob{\sigma,\tau''}{\delta}{K_N=l\text{ and } \forall 1\leq m\leq N, K_m\not\in\targets}>0\enspace.
\ee
By definition of $\tau''$, there is positive probability that $\tau''$ picks randomly the support $B\in\LL$
and switches to $\tau$ with initial support $B$.
By definition, $\tau$ is positively winning from $B$ hence there exists $l\in B$ such that:
\[
\forall \sigma', \prob{\sigma',\tau}{l}{\forall n\geq 1, K_n \not\in\targets} >0\enspace,
\]
together with~\eqref{eq:BBB} it proves
$\prob{\sigma,\tau''}{\delta}{K_N=l\text{ and } \forall m\geq 1, K_m\not\in\targets}>0$
hence
~\eqref{eq:taupp}.

%

\medskip

Now we consider the case where $\LL_p$ is empty.
According to Proposition~\ref{prop:LLgames},
player $1$ has a pessimistic belief strategy $\sigma$
which is almost-surely winning in the $\LL$-game
from every support $L\not\in\LL$.
This ensures, for every $\delta$ whose support is $L\in\LL$,
for every strategy $\tau$,
\be\label{eq:pbelinfty}
\prob{\sigma,\tau}{\delta}{\forall n\geq 1, \pbelun(L,C_1,\ldots, C_n)\not\in\LL\text{ or } \exists m\leq n, K_m\in\targets}=1\enspace.
\ee

We start with proving for each $L\not\in\LL$
there exists $N_L\in\NN$ such that for every strategy $\tau$,
for every distribution $\delta$ with support $L$,
\begin{equation}\label{eq:unif3}
\prob{\sigma,\tau}{\delta}{
\exists n\leq N_L, K_n\in\targets
}\geq \frac{1}{2}\enspace.
\end{equation}
We suppose such an $N_L$ does not exist and seek for a contradiction.
Suppose for every $N$ there exists $\tau_{N}$ and $\delta_N$ such that~\eqref{eq:unif3} does not hold.
We can suppose $\tau_N$ is deterministic i.e. $\tau_N:\signauxdeux^* \to \actionsdeux$,
and $\delta_N$ converges to some distribution $\delta$, whose support is included in $L$.
Using Koenig's lemma, it is easy to build a strategy $\tau:\signauxdeux^*\to \actionsdeux$ such that for
infinitely many $N$,
\[
\prob{\sigma,\tau}{\delta_N}{
\exists n\leq N, K_n\in\targets
}\leq \frac{1}{2}\enspace.
\]
Taking the limit when $N\to\infty$, we get: 
\[
\prob{\sigma,\tau}{\delta}{
\exists n\geq 1, K_n\in\targets
}\leq \frac{1}{2}\enspace,
\]
which
contradicts the fact that $\sigma$ is almost-surely winning from $L$,
since the support of $\delta$ is included in $L$.
This proves the existence of $N_L$ such that~\eqref{eq:unif3} holds.

Now let $N=\max\{N_L\mid L\not\in\LL\}$ and let $\sigma'$ be the pessimistic belief strategy for player $1$
similar to $\sigma$, except every $N$ steps the memory is reset,
formally:
$\sigma'(L)(c_1,\ldots,c_n) = \sigma(\pbelun(L,c_1,\ldots,c_{(n /N)*N}))(c_{(n / N)*N}\cdots c_n)$.
Then whatever be the strategy played by player $2$,
according to~\eqref{eq:pbelinfty} as long as a target state is not reached,
the memory of $\sigma'$ will stay outside $\LL$.
Then according to~\eqref{eq:unif3}, when playing $\sigma'$,
every $N$ steps there is probability at least $\frac{1}{2}$ to reach a target state,
knowing that it was not reached before,
hence there is probability $0$ of never reaching a target state.
Consequently, $\sigma$ is almost-surely winning from any support $L\not\in\LL$.
\end{proof}

Now we can prove Theorem~\ref{theo:twoplayers2}.
\begin{proof}[Proof of Theorem~\ref{theo:twoplayers2}]
Let $\LL_0,\LL_1,\ldots$ be the sequence defined by
$\LL_0=\emptyset$ and for every $n\in\NN, \LL_{n+1}\subseteq\parties{\states}$ is the set of supports
positively winning for player $2$ in the $\LL_n$-game.
Then $\LL_0\subseteq \LL_1\subseteq \ldots$ and $\parties{\parties{\states}}$
is finite hence there is a limit $\LL_\infty$ to this sequence.

Every $\LL_n$ is upward-closed hence according to Proposition~\ref{prop:upward},
every support in $\LL_\infty$ is positively winning for player $2$.
Moreover, according to Proposition~\ref{prop:LLgames},
player $2$ has a positively winning strategy with finite memory whose memory states are sets of pairs a state and a
pessimistic belief of player $1$.

By definition of $\LL_\infty$,
the only support positively winning in the $\LL_\infty$-game are in $\LL_\infty$.
Hence according to Proposition~\ref{prop:upward} again,
every support not in $\LL_\infty$ is almost-surely winning for player $1$.
Moreover, according to Proposition~\ref{prop:LLgames},
player $1$ has a pessimistic belief almost-surely winning strategy.

The computation of $\LL_\infty$ can be achieved in doubly-exponential time,
because according to Proposition~\ref{prop:LLgames} each step can be carried on
in time doubly exponential in $\states$ and since the sequence $(\LL_n)_{n\in\NN}$
is monotonic its length is at most exponential in $|\states|$.
\end{proof}

\section*{Conclusion}
We considered stochastic reachability games with partial observation on both sides.
We established a determinacy result: such a game is either almost-surely
winning for player $1$, surely winning for player $2$ or positively winning for both players.
Despite its simplicity, this result is not so easy to prove.
Also we gave algorithms for deciding in doubly-exponential time which of the three cases hold.

A natural question is whether these results extend are true for B{\"u}chi games as well?
The answer is "partially".

One one hand, it is possible to prove that a game is either almost-surely winning for player
$1$ or positively winning for player $2$ and to decide in doubly-exponential time which of the two cases hold.
This can be done by techniques almost identical to the ones in this paper.

On the other hand, it was shown recently that the question "does player $1$ has a
\emph{deterministic} strategy for winning positively a B{\"u}chi game?" is undecidable~\cite{bbg},
even when player $1$ receives no signals and player $2$ has only one action.
It is quite easy to see that "deterministic" can be removed from this question,
without changing its answer. Hence the only hope for solving positive winning
for B{\"u}chi games is to consider subclasses of partial observation games
where the undecidability result fails, an interesting question.

\bibliographystyle{alpha}
\bibliography{signalas}

\end{document}